\documentclass[aps,prl,twocolumn,showpacs,superscriptaddress,groupedaddress]{revtex4}  
\usepackage{graphicx}  
\usepackage{dcolumn}   
\usepackage{bm}        
\usepackage{amssymb}   
\usepackage{gensymb}
\usepackage{amsmath}

\usepackage{color}
\usepackage[normalem]{ulem}

\def\={\,=\,}

\begin{document}

\title{Supercurrent Flow in Multi-Terminal Graphene Josephson Junctions
}

\affiliation{Department of Physics, Duke University, Durham, NC 27708, USA}
\affiliation{Department of Physics and Astronomy, Appalachian State University, Boone, NC 28607, USA}
\affiliation{Advanced Materials Laboratory, NIMS, Tsukuba 305-0044, Japan}
\affiliation{Department of Physics, City University of Hong Kong, Kowloon, Hong Kong SAR}
\author{Anne W.~Draelos} \affiliation{Department of Physics, Duke University, Durham, NC 27708, USA}
\author{Ming-Tso~Wei} \affiliation{Department of Physics, Duke University, Durham, NC 27708, USA}
\author{Andrew~Seredinski} \affiliation{Department of Physics, Duke University, Durham, NC 27708, USA}
\author{Hengming~Li} \affiliation{Department of Physics and Astronomy, Appalachian State University, Boone, NC 28607, USA}
\author{Yash~Mehta} \affiliation{Department of Physics and Astronomy, Appalachian State University, Boone, NC 28607, USA}
\author{Kenji~Watanabe} \affiliation{Advanced Materials Laboratory, NIMS, Tsukuba 305-0044, Japan}
\author{Takashi~Taniguchi} \affiliation{Advanced Materials Laboratory, NIMS, Tsukuba 305-0044, Japan}
\author{Ivan V. Borzenets} \affiliation{Department of Physics, City University of Hong Kong, Kowloon, Hong Kong SAR}
\email{iborzene@cityu.edu.hk}
\author{Fran{\c c}ois~Amet} \affiliation{Department of Physics and Astronomy, Appalachian State University, Boone, NC 28607, USA}
\author{Gleb~Finkelstein} \affiliation{Department of Physics, Duke University, Durham, NC 27708, USA}
\email{gleb@phy.duke.edu}
     
\date{\today}

\begin{abstract}
We investigate the electronic properties of ballistic planar Josephson junctions with multiple superconducting terminals. Our devices consist of monolayer graphene encapsulated in boron nitride with molybdenum-rhenium contacts. Resistance measurements yield multiple resonant features, which are attributed to supercurrent flow among adjacent and non-adjacent Josephson junctions. In particular, we find that superconducting and dissipative currents coexist within the same region of graphene. We show that the presence of dissipative currents primarily results in electron heating and estimate the associated temperature rise. We find that the electrons in encapsulated graphene are efficiently cooled through the electron-phonon coupling.
\end{abstract}

\maketitle

Superconducting proximity effects in graphene-based superconductor-normal-superconductor (SNS) devices attracted researchers' attention early on \cite{Heersche,Miao,Du}. Ultra-clean suspended graphene and heterostructures of graphene encapsulated in boron nitride further enable the study of proximity-induced superconductivity in the ballistic regime \cite{Calado,Shalom,Allen,Kumaravadivel,Borzenets}. While these and following works investigated supercurrent flow in two-terminal junctions, here we are interested in more complex SNS structures where a normal metal is proximitized by several superconducting electrodes. 

In such devices, the interplay of supercurrent flow involving different pairs of superconducting contacts could be highly non-trivial. For example, in a simple four-terminal configuration, biasing one terminal with respect to an adjacent terminal has been predicted to induce a ``phase drag'' between the other two terminals \cite{Omelyanchouk,Amin}. Furthermore, the energy spectrum of a multi-terminal SNS device can emulate a band structure, where the superconducting phases play the role of quasi-momenta of a ``crystal'' of arbitrary dimension \cite{Mai,Xie,Eriksson,Riwar}. This effective band structure can accommodate Weyl points, and multi-terminal Josephson junctions are therefore expected to provide insights into the physics of topological materials. On the experimental side, topological properties of multi-terminal structures have been investigated in Refs.~\cite{Strambini,Vischi}. Elsewhere, it has been reported that complex spectra of Andreev reflections could originate from non-local entanglement of two or more Cooper pairs in three or more leads \cite{Freyn,Jonckheere,Pfeffer,Cohen}. 


Encapsulated graphene is especially promising for the investigation of multi-terminal SNS devices. Indeed, clean encapsulated samples are ballistic and could enable efficient Josephson coupling between superconducting contacts separated by several microns \cite{Borzenets}. Here we report on the first observation of complex supercurrent flow in a four-terminal graphene SNS device. Our findings include a highly unusual coexistence of superconducting and dissipative currents in the same physical location.

\begin{figure*}
\includegraphics[width=7in]{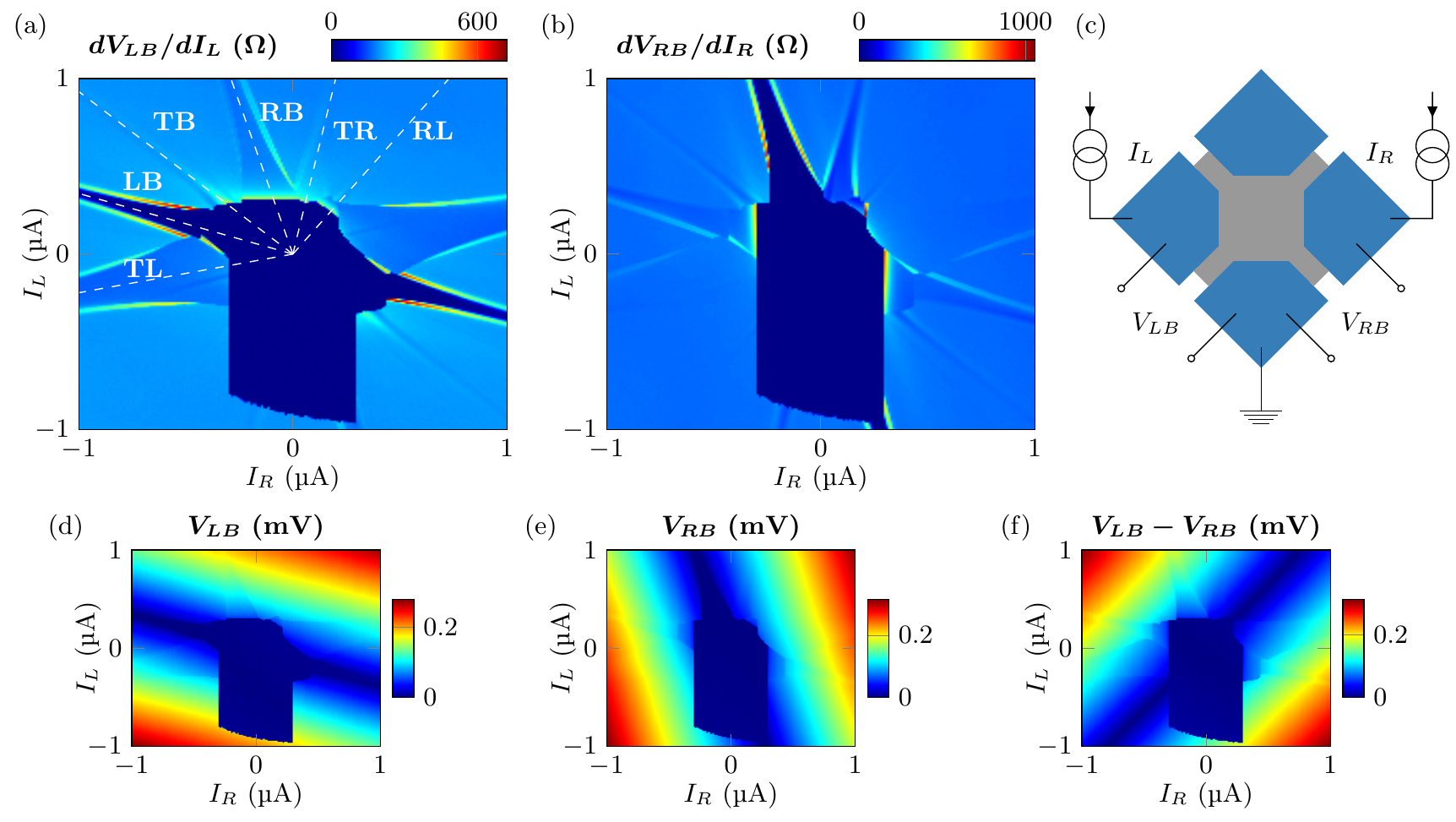}
\caption{\label{fig:Multiplet} 
(a,b) Differential resistances $dV_{LB} / dI_L$ and $dV_{RB} / dI_R$, measured at $V_G$ = 5 V in the set-up shown in panel (c). The darkest region in the center of both figures corresponds to supercurrent across the whole sample. Maps were obtained by setting a single value of $I_R$ and sweeping $I_L$ from positive to negative values. The resulting superconducting region in the center of the maps is asymmetric in shape and extends further to negative $I_L$ due to the hysteresis between the retrapping and switching transitions (into and out of the superconducting state). Arm-shaped dark regions of low resistance extend from the central rectangle (dashed lines); we show that they correspond to supercurrent flowing only between some pairs of terminals, as labeled.
(c) Diagram of the sample with four superconducting terminals (blue) contacting a shared graphene region (grey). In the primary measurement configuration, currents $I_L$ and $I_R$ are sourced from the left and right contacts to the grounded bottom contact. Voltage difference between any two terminals could then be measured, such as e.g. $V_{LB}$ and $V_{RB}$, the left-bottom and right-bottom voltage difference. 
(d,e) DC voltage maps $|V_{LB}|$ and $|V_{RB}|$ measured concurrently with the differential resistances maps in (a,b), respectively. It is clear that the orientation of the lines $V_{LB}=0$ and $V_{RB}=0$ (darkest blue) coincides with the major features in the differential resistance maps (a) and (b). 
(f) A map of the voltage difference between the left and the right contacts, $|V_{LR}|$. The line of equal voltage $V_{LR}=0$ runs diagonally from bottom left to top right, and a faint feature at the same location is also visible in maps (a,b). }
\end{figure*}

The fabrication of these devices was detailed in prior works, which showed ballistic supercurrents in Josephson junctions up to two microns long \cite{Amet, Borzenets}. Exfoliated monolayer graphene is encapsulated with hexagonal boron nitride using a standard stamping method \cite{Wang}. The resulting stack is placed onto a p-doped silicon substrate with a 300 nm oxide layer. Carrier density in the graphene is controlled by a back gate voltage $V_G$. After an anneal in an argon/oxygen mixture, the stack is patterned with electron-beam lithography and then cut by reactive-ion etching to form the desired square mesa (3 $\mu$m x 3 $\mu$m). Contact electrodes are patterned on each corner, followed by a self-aligned etch and deposition of molybdenum-rhenium. The electrodes are $\sim 100$ nm thick, have a critical field of at least 9 T, a critical temperature $T_C \sim 10$ K, and a measured superconducting gap $\Delta \sim$ 1.2 meV \cite{Calado,Amet}. Each contact is 500 nm away from its adjacent neighbor at the closest point, and is $\sim$ 2 $\mu$m away from the contact diagonally across. 

The samples were cooled in a Leiden Cryogenics dilution refrigerator to a base temperature of $\sim 45$ mK. Figure \ref{fig:Multiplet}c shows a schematic of the transport measurement setup: two current sources were used, each combining a small AC excitation current of 10 nA and a DC bias of up to $\sim 1 \mu$A (sufficient to switch a junction from the superconducting to the normal state). The bottom contact was grounded, and the top terminal remained floating, with no net current flowing through it. Two lock-in amplifiers synchronized to separate frequencies were used when measuring the differential resistance across various junctions. The DC voltage across a junction was concurrently measured, using a multichannel digital acquisition system (NI-6363 DAQ). 

\begin{figure}[tbp]
\begin{center}
\includegraphics{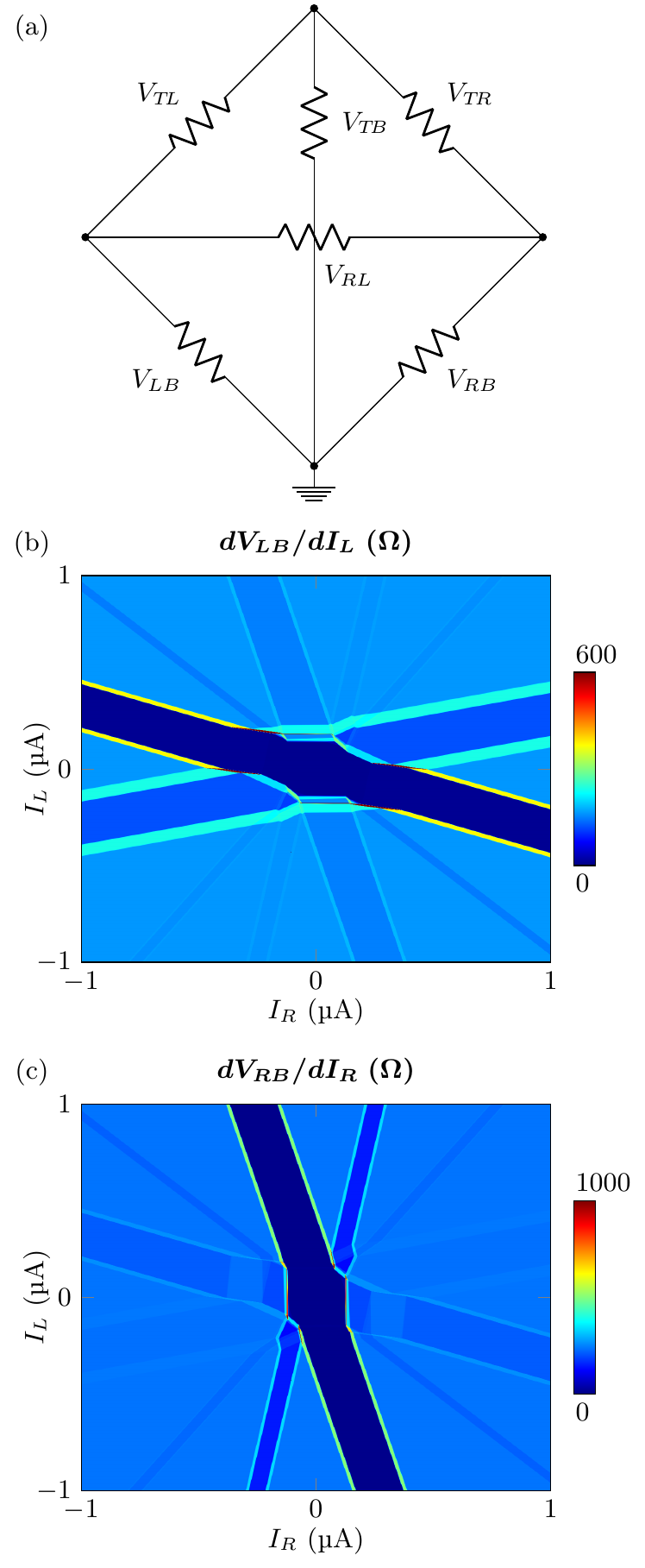}
\end{center}
\caption[Identification of supercurrent lines in a multi-terminal device]{
(a) An equivalent circuit diagram that represents the coupling between each pair of terminals as a resistor. Each resistor could be replaced with an effective short to indicate a superconducting coupling between the corresponding pair of terminals, and consequently the locally measured voltage would be zero. 
(b, c) Maps of differential resistance for the left bottom and right bottom junctions, respectively, generated from a numerical simulation. The six major features in Fig.~\ref{fig:Multiplet}a,b are realistically reproduced in both maps, with lines of equal voltage appearing at certain orientations within the $I_L$-$I_R$ plane.}
\label{fig:Model}  
\end{figure}

Figure \ref{fig:Multiplet}a,b shows differential resistance maps of the left-bottom (LB) and right-bottom (RB) junctions as current is applied from both the left ($I_L$) and right ($I_R$) leads to the grounded bottom lead. In this way, the maps of the differential resistance $\frac{dV_{LB}}{dI_L}$ and $\frac{dV_{RB}}{dI_R}$ can be simultaneously measured as a function of both currents. 
A central dark region of zero resistance in both maps indicates supercurrent through the whole sample. Due to the difference between the switching and the retrapping currents, this region extends to the bottom part of the maps, which are measured with $I_R$ held fixed while $I_L$ is swept from positive to negative bias.

Beyond the central roughly rectangular feature, this superconducting region is additionally extended into several narrow diagonal ``arms'' of suppressed resistance. To help identify the origin of these arms, we plot the DC voltage across the left and right junctions $V_{LB}$ and $V_{RB}$.  Figure~\ref{fig:Multiplet}d shows a line of zero voltage drop across the left-bottom junction ($V_{LB} = 0$), which naturally corresponds to the dominant arm of suppressed resistance in the resistance of the left-bottom junction (Figure~\ref{fig:Multiplet}a). Similarly, for the right bottom junction the dominant arm in Figure~\ref{fig:Multiplet}b is due to a zero voltage drop $V_{RB} = 0$ (Figure~\ref{fig:Multiplet}e). Both features result in resistance being suppressed to zero by the supercurrent flowing between the two contacts being measured. We furthermore find a fainter version of each feature in the complementary map: the LB pair turning superconducting affects the measured RB resistance and vice versa. Beyond these obvious supercurrent branches, we now show that the other arms found in Figure~\ref{fig:Multiplet}a,b correspond to superconducting coupling between other pairs of terminals. 

Specifically, we find that these additional features are due to the incorporation of the top, ungrounded and otherwise unused terminal. Keeping the sources and ground terminals the same and moving the voltage probes to the top left and right junctions allows us to identify the remaining wide features in Figure~\ref{fig:Multiplet}b,c as the lines of equal DC voltage $V_{TL}=0$ and $V_{TR}=0$. This condition indicates that the pairs of terminals top-left or top-right are coupled by a superconducting current. The relative strength of these features is also understandable: the LB resistance $\frac{dV_{LB}}{dI_L}$ is strongly affected by the supercurrent in the adjacent TL junction with which it shares the left terminal (Figure~\ref{fig:Multiplet}b). On the other hand, the supercurrent in the TR junction has relatively less effect on LB because no terminals are directly shared, and the corresponding feature is faint. Instead, the supercurrent in the TR junction creates a prominent feature in the $\frac{dV_{RB}}{dI_R}$ map in Figure~\ref{fig:Multiplet}c, as the two junctions share the right terminal. 


The maps also show two faint and narrow features labeled RL and TB, which we attribute to a superconducting coupling between diagonally opposite contacts. Indeed, the location of the RL line coincides with the condition $V_{RL}=0$ visible in Figure~\ref{fig:Multiplet}f, which shows the $V_{RL}$ map. It should be noted that the induced coherence length ($\xi \approx$ 500 nm) is comparable to the distance between the neighboring superconducting contacts and smaller than the diagonal distances R-L and T-B \cite{Borzenets}. Therefore, the supercurrents across the diagonals should be several times weaker than the supercurrents coupling the neighboring contacts. Nonetheless, the features corresponding to the supercurrent flowing along the sample diagonals are clearly observed. Furthermore, we surprisingly find areas of the maps in which only one pair of diagonally opposite contacts is coupled by a supercurrent, while the supercurrent between any other pair of contacts is suppressed. This means that a supercurrent could flow along one of the sample diagonals, while dissipative currents flow along the opposite diagonal and between any neighboring contacts.

\begin{figure*}[tbp]
\begin{center}
\includegraphics[width=7in]{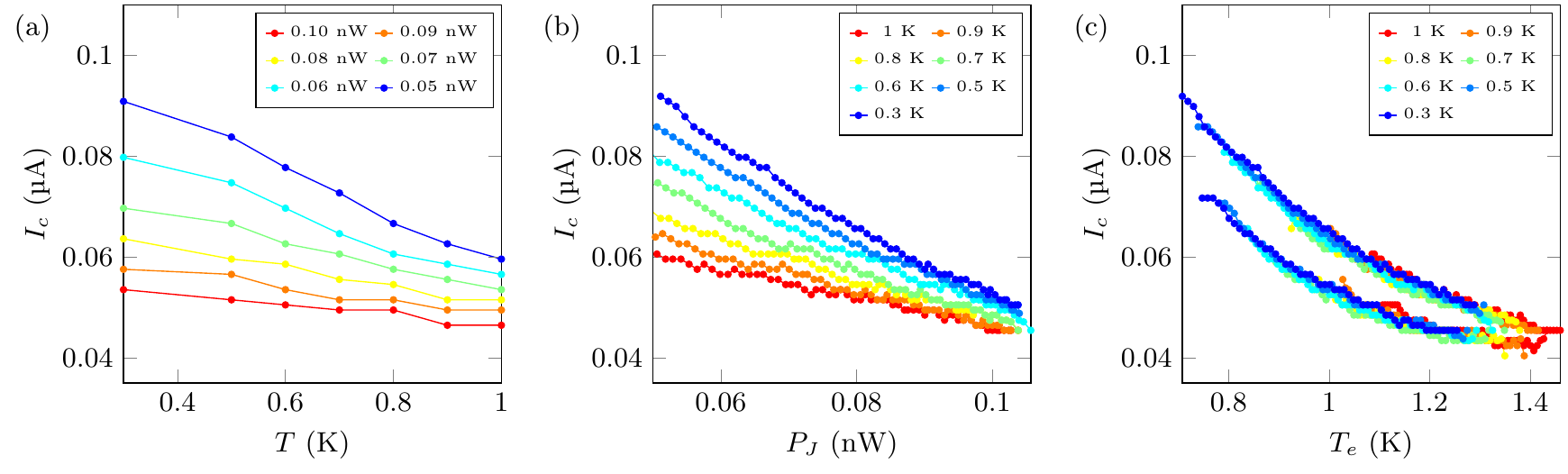}
\end{center}
\caption[Heating effects on the magnitude of the supercurrent]{(a,b) Critical current in the LB junction plotted as a function of the substrate temperature and heating power, respectively. For a given substrate temperature, we extract the dependence of the critical current $I_C$ (measured along the $I_L$ axis) on the heating current $I_R$. The heating power $P_J$ is calculated as the product of $I_R$ and $V_R$ (measured separately). (c) The critical current in the two junctions, plotted as a function of electron temperature $T_{e}$. $T_{e}$ is calculated from the equilibrium between Joule heating and electron-phonon cooling as $(T^{3}+P_{J}/\Sigma)^{\frac{1}{3}}$. Importantly, the same $\Sigma$ is used for both junctions. The observed universal behavior shows how the decay of each superconducting branch can be attributed solely to electron heating.}
\label{fig:Heating}  
\end{figure*}

To quantitatively understand the pattern of supercurrent flow, we examine a circuit diagram consisting of six Josephson junctions connecting the four terminals (Figure~\ref{fig:Model}a). Away from the central superconducting region of the map in Figure~\ref{fig:Multiplet}a,b, the diagonal arms of suppressed resistance correspond to one of the junctions turning superconducting (i.e. its resistance in the network model is switched to zero), while the rest remain resistive. By measuring $\frac{dV_{LB}}{dI_L}$ and  $\frac{dV_{RB}}{dI_R}$ when the system is fully resistive, and when some of the junctions are in the superconducting state, we can construct a system of equations allowing us to solve for the normal resistances of all junctions. We find that the neighboring leads are effectively coupled by  $\sim 300 \Omega$, while the diagonal resistances are $\sim 1.5 - 2 k\Omega$.

Once the normal state resistances are determined, and the critical currents $I_C$ of each junction are estimated from the maps, we can model the current flow in the system ~\cite{note1}. The network includes six resistively shunted junctions (RSJ), each consisting of a normal resistor (Figure~\ref{fig:Model}a) in parallel with a junction of critical current $I_C$. Figures \ref{fig:Model}b,c show the resultant maps of differential resistance for the left and right junctions, corresponding to the experimental maps in Figure~\ref{fig:Multiplet}a,b. The model captures the overall features of the experiment, such as the existence of the central superconducting region and the extending diagonal ``arms''. The values of the resistance in the arms, as well as the slopes of the arms, are constrained by our extracted values of the network resistors. Note, that in our model the diagonal junctions TB and RL do not directly interact, even though they share the same physical space. This assumes that the coexisting dissipative and superconducting currents do not strongly interfere with each other.

While all six of the supercurrent arms are reproduced in this model, the decay of $I_C$ at higher bias currents is not, because the model does not account for the effects of dissipation within the device. Indeed, it has been previously shown that the critical current of an SNS structure is strongly affected by self-heating \cite{Courtois}. Some of us have explored this effect in graphene-based structures \cite{Borzenetsbottleneck}. To quantify the effects of Joule heating in the present device, we measure similar maps of differential resistance at different substrate temperatures. From such maps, we extract the value of the critical current in the measured junction as a function of the substrate temperature $T$ (Figure ~\ref{fig:Heating}a). The different curves correspond to different applied heating power $P_J$, calculated as $I_R V_R$ to represent the Joule heating due to the right current when the left junction is superconducting. Figure~\ref{fig:Heating}b represents the same data plotted vs. $P_J$ at different $T$. Clearly, $I_C$ is suppressed by both $P_J$ and $T$. 

While the dissipative currents heat the electron system, the superconducting gap in the contacts prevents the efficient cooling by outflow of hot electrons \cite{Borzenetsbottleneck}. As a result, heat can only dissipate via the relatively inefficient electron-phonon coupling. The power dissipated through electron-phonon cooling scales like $P_{e-ph}=\Sigma (T^{\delta}_{e}-T^{\delta})$, where $\Sigma$ is the electron phonon coupling constant integrated over the area of the device, $T_e$ is the temperature of the electron system, and $T$ is the phonon bath temperature, assumed to be unaffected by the dissipated heat and equal to the substrate temperature. The exponent $\delta$ could be 3 or 4 depending on the temperature range, and the mean free path; for a recent summary of the literature, see Ref.~\cite{Prober}. If all the Joule heat is dissipated via phonons, the electron temperature at equilibrium is expected to reach $T_{e}=(T^{\delta}+P_J/\Sigma)^{1/\delta}$. Using a single fitting parameter $\Sigma$, we can show that the critical current $I_C$ in both the LB and RB junctions scales universally as a function of the equilibrium electron temperature, $T_{e}=(T^{3}+P_J/\Sigma)^{\frac{1}{3}}$ (Figure~\ref{fig:Heating}e). Here, we used $\delta =3$; we have found that taking $\delta=4$ does not allow us to successfully scale the data. $\Sigma$ is found to be 30 pW/K$^3$ (or scaled per graphene area $\sim10$ W/m$^2$K$^3$), which is significantly larger than the results found in non-encapsulated graphene \cite{Prober}. We speculate that the more efficient cooling in our case could be explained either by scattering at the edges of the graphene crystal which could momentum relaxation, or by direct emission of phonons into the BN substrate, both of which which enhance electron-phonon scattering.\cite{Kong}.

The presence of additional dissipative current in an SNS structure should affect the superconducting properties of the junction beyond simple heating \cite{Morpurgo,Klapwijk1, Klapwijk2, Giazotto}. Such effects have been previously observed in diffusive metallic SNS junctions, where injection of a dissipative current resulted in reversing the direction of the supercurrent. However, here no such influence was observed. This lack of notable effects is either due to the ballistic nature of our device, or likely due to the limitations of the measurement system. Additional measurements would be required to search for any change of the current-phase relation in our samples.

In summary, we have experimentally studied the complex supercurrent flow in a four-terminal graphene SNS junction. We explained the six distinct branches that appear in the resistance maps. Interestingly, we have observed the regime where supercurrent and dissipative current coexist in the same area of graphene. The results could be successfully modeled with a network of six shunted Josephson junctions. Finally the effects of the dissipative current on the magnitude of the supercurrent can be accounted for by self-heating in the device. Our experiment opens the prospects for future search of topological effects in multi-terminal graphene devices \cite{Mai,Xie,Eriksson,Riwar,Strambini,Vischi}.

A.W.D., and G.F. thank the  Division of Materials Sciences and Engineering, Office of Basic Energy Sciences, U.S. Department of Energy, under Award DE-SC0002765. M.T.W. and A.S. thank ARO Award W911NF-16-1-0122. F.A. acknowledges the ARO under Award W911NF-16-1-0132. I.V.B. acknowledges CityU New Research Initiatives/Infrastructure Support from Central (APRC): 9610395. This work was performed in part at the Duke University Shared Materials Instrumentation Facility (SMIF), a member of the North Carolina Research Triangle Nanotechnology Network (RTNN), which is supported by the National Science Foundation (Grant ECCS-1542015) as part of the National Nanotechnology Coordinated Infrastructure (NNCI). K. W. and T. T. acknowledge support from JSPS KAKENHI Grant No. JP15K21722 and the Elemental Strategy Initiative conducted by the MEXT, Japan. T. T. acknowledges support from JSPS Grant-in-Aid for Scientific Research A (No. 26248061) and JSPS Innovative Areas `Nano Informatics' (No. 25106006).

\end{document}